
%
%
\magnification=1200
\def\setup{\count90=0 \count80=0 \count91=0 \count85=0
\countdef\refno=80
\countdef\secno=85}
\def\R{\vrule height5.85pt depth.2pt \kern-.05pt \tt R}
\def\Box{\vbox{\hrule
                    \hbox{\vrule height 6pt \kern 6pt \vrule height 6pt}
                    \hrule}\kern 2pt}
\def\lo{\raise2pt\hbox{$<$}\kern-7pt\raise-2pt\hbox{$\sim$}}
\def\go{\raise2pt\hbox{$>$}\kern-7pt\raise-2pt\hbox{$\sim$}}

\def\vline{{\vrule height8pt depth4pt}\; }
\def\Vline{{\vrule height13pt depth8pt}\; }
\def\sub#1{{\lower 8pt \hbox{$#1$}}}

\def\Del{{\raise.5ex\hbox{$\bigtriangledown$}}}
\def\DEL#1{{\raise.5ex\hbox{$\bigtriangledown$}\raise 8pt \hbox{\kern -10pt
                 \hbox{$#1$}} }}
\def\autoeq{ {\global\advance\count90 by1} \eqno(\the\count90) }
\def\autoeql{ {\global\advance\count90 by1} & (\the\count90) }
\def\autosec{ {\global\advance\secno by 1} (\the\secno) }
\def\e{\hbox{e}}
\def\Lie#1{{\cal L}{\kern -6pt
            \hbox{\raise 1pt\hbox{-}}\kern 1pt} _{\vec{#1}}}

\def\Z{Z \kern-5pt \hbox{\raise 1pt\hbox{-}}\kern 1pt}
\def\eye{\bigcirc\kern -7pt\bigcirc}
\def\autoref{ {\global\advance\refno by 1} \kern -5pt [\the\refno]\kern 2pt}
\def\readref#1{{\count100=0 \openin1=refs\loop\ifnum\count100
<#1 \advance\count100 by1 \global\read1 to \title \global\read1 to \author
\global\read1 to \pub \repeat\closein1  }}
\def\reftitle{{ \kern -3pt \vtop{ \hbox{\title} \hbox{\author\ \pub} } }}
\def\ref{{ \kern -3pt\author\ \pub \kern -3.5pt }}
\setup
\centerline{  }
\line{\hfill \hbox{DIAS-STP-94-21}}
\line{\hfill \hbox{June 10th 1994} }
\vskip 2cm
\centerline{\bf Symplectic Geometry And Hamiltonian Flow Of}
\centerline{\bf The Renormalisation Group Equation}
\vskip 1.2cm
\centerline{Brian P. Dolan}
\vskip .5cm
\centerline{\it Department of Mathematical Physics, St. Patrick's College}
\centerline{\it Maynooth, Ireland}
\centerline{and}
\centerline{\it Dublin Institute for Advanced
Studies}
\centerline{\it 10, Burlington Rd., Dublin, Ireland}
\vskip .5cm
\centerline{e-mail:bdolan@maths.may.ie}
\vskip 1.5cm
\centerline{ABSTRACT}
\noindent It is argued that renormalisation group flow
can be interpreted as being a Hamiltonian vector flow on a phase
space which consists of the couplings of the theory and their conjugate
\lq\lq momenta", which are the vacuum expectation values of the corresponding
composite operators. The Hamiltonian is linear in the conjugate
variables and can be identified with
the vacuum expectation value
of the trace of the energy-momentum operator. For theories
with massive couplings the identity operator plays a central role
and its associated coupling gives rise to
a potential in the flow equations.  The evolution of
any quantity , such as $N$-point Green functions, under
renormalisation group flow can be obtained from its
Poisson bracket with the Hamiltonian. Ward identities can
be represented as constants of the motion which act as
symmetry generators on the phase space via the Poisson bracket
structure.
\vskip 1cm

\vfill\eject
{\bf \S 1 Introduction \hfill}
\vskip .5cm
The history of relativistic quantum field theory is plagued with
divergences. The canonical cure for this is to replace the
(infinite) bare co-ordinates $g_0^a$ by (finite) renormalised
co-ordintates $g_R^a(\kappa)$ at some renormalisation point $\kappa$.
This requires regulating the divergences and choosing a subtraction
prescription by introducing counter terms.
Even for theories with no divergences it is sometimes useful
to introduce counter terms and define renormalised couplings -
an example of this is the $\epsilon$-expansion around the Wilson-Fisher
non-trivial fixed point in $3$-dimensional massive $\varphi^4$
theory
\autoref\newcount\Brezin\Brezin=\refno

However knowledge of the renormalised
couplings alone is not sufficient, one must also know what
the counter terms are. This is (almost) equivalent to knowing
the $\beta$-functions of the theory, since the counter terms
are specified by the difference
$\Delta g^a(\kappa)=g_0^a-g_R^a(\kappa)$. Demanding that the bare
couplings are independent of the renormalisation point gives
$$\beta^a={dg_R^a\over dt}=-{d(\Delta g^a)\over dt},\autoeq$$
\newcount\befunc\befunc=\count90
where $t=\ln\kappa$.

The traditional approach has been first to choose the
counter terms and then calculate the $\beta$-functions.
However the counter terms contain a certain ambiguity, they can be modified
by adding a finite function of the couplings to $\Delta g^a$.
Thus, as is well known, the $\beta$-functions are not unique,
they depend on the subtraction procedure.
In practice though they are constrained by the requirement
that a renormalisation prescription be chosen which leads to a
perturbation theory that makes sense i.e. converges
reasonably quickly, at least asymptotically. Of course, the
$\beta$-functions
are not completely arbitrary - they have zeros which cannot be
removed by changing prescription. Viewing them as
vector fields (\lq\lq velocities") on the space of couplings,
a zero of the vector field in one prescription remains a zero
in any other prescription (of course individual
components of the vector $\vec{\beta}$ may vanish in one
prescription and not in another, but a zero of the vector
field requires the entire
vector to vanish). Thus, in one sense at least, a change in
renormalisation prescription can be thought of as a co-ordinate
transformation (diffeomorphism) on the space of couplings since
this also leaves the zeros of a vector field unchanged
(although it may, and in general would,
change the numerical values of the co-ordinates
of the point at which the zero occurs).

One can imagine turning the logic round and first choosing
the $\beta$-functions and then using equation (\the\befunc)
to determine the counter terms (up to an arbitrary constant
i.e. a renormalisation group invariant).
Of course the choice must be judicious - it would be crazy to
choose a positive $\beta$-function for massless QCD, the
resulting theory would be completely unstable and probably would
not even exist.
In theories
with more than one coupling, however, there is more freedom.
If one were unlucky enough to choose a $\beta$-function which
pointed away from a nearby attractive fixed point the RG evolution
would presumably force it to turn round and point
in towards the fixed point - behaviour which cannot happen if there
is only one coupling. Choosing a $\beta$-function with the
\lq\lq wrong sign" in a theory with more than one coupling
is not pathological, it is merely an indication of a \lq\lq bad" choice
of co-ordinates. Transforming to \lq\lq good" co-ordinates
(i.e. co-ordinates which have a sensible physical
interpretation and are not just abstract parameters) would require
performing a diffeomorphism which \lq\lq unwinds" the RG trajectory
so that it flows into the fixed point in a straight line.
In actual fact, from a topological point of view, the notion
of \lq\lq pointing away" does not make sense
unless a metric is defined on the
space of couplings. To have a concept of \lq\lq pointing away"
requires having a definition of angles, and this needs a metric.
Without a metric any direction is much the same as any other,
it is difficult to navigate in a space without a metric!

Another quantity which should not change under a co-ordinate
transformation is the signature of the matrix of anomalous
dimensions, $\partial_a\beta^b $, at a fixed point. The number
of positive and negative eigenvalues of this matrix determines
in how many directions the RG flow is attracted to the
fixed point and in how many it is repelled from it. For massless QCD
this matrix is one dimensional and this is another way to see
that the sign of the $\beta$-function cannot be changed
by a co-ordinate transformation in a
theory with only one coupling.

There is an analogy here with classical mechanics. Consider
a theory with $n-1$ couplings, $g^a,\;a=1,\ldots,n-1$ (from now on the
subscript $R$ on couplings will be dropped in this section
- all couplings are renormalised
unless otherewise indicated).
Denote the space of couplings by $\cal M$. The $\beta$-functions
$\beta^a(g)$ constitute a vector field on the $n-1$-dimensional
differentiable manifold $\cal M$ (for the moment the
topology and global properties of $\cal M$ will not be
relevant and it may as well be taken to be ${\bf R}^{n-1}$ -
consideration of the global topology will be restricted to a few
comments in the final section).
The $2n-2$ dimensional space with co-ordinates $(g^a,\beta^a)$
(the tangent bundle $T({\cal M})$) is thus analogous to
the configuration space of co-ordinates and velocities in classical
mechanics. Choose a point in $T({\cal M})$ and the evolution of the
system is determined by the dynamics. All the necessary
information for computing the RG evolution is contained in the
generating functional (or free energy in statistical mechanics)
$W(g,t)=\int w(g,t)d^Dx=-\ln Z$
where $w(g,t)$ is the free energy density.
\footnote*{For simplicity we
shall work on flat $D$-dimensional Euclidean space, so that
translational invariance ensures that $w(g,t)$ is independent
of position, but the concepts presented here are more general
than this.}
Just as in classical mechanics one can imagine
transforming to a phase space, with co-ordinates $(g^a,\phi_a)$
(the cotangent bundle $T^*({\cal M})$) where $\phi_a$ are
\lq\lq momenta" dual to the \lq\lq velocities" $\beta^a$.
A natural choice for the $\phi_a$ is the vacuum expectation
value of the (in general composite) operator associated with the
coupling $g^a$, $\phi_a={\partial w(g,t)\over\partial g^a}$
\autoref\newcount\DenjoeChris\DenjoeChris=\refno.
In order to streamline some of the formula
it will be convenient to re-scale all the couplings by their canonical
dimensions so that they are dimensionless.
When this is done the variables $\phi_a$ are densities
with canonical mass dimension $D$.

Following O'Connor and Stephens [\the\DenjoeChris]
one can ask if there might be some notion
of a Hamiltonian function on phase space $H(g,\phi,t)$ which could
govern the RG evolution of the couplings and the expectation
values together so that the RG flow can be regarded as a Hamiltonian
vector flow on phase space. The answer is yes and the construction
turns out to be remarkably simple. A Hamiltonian will be
presented in section three which is {\it linear} in the momenta, rather
than quadratic as in non-relativistic particle mechanics.
It is in fact minus the expectation value of the
trace of the energy-momentum operator, $H=-<T>$.
Despite
the linearity of the Hamiltonian in the momenta,
the dynamical evolution is
not trivial and some examples will be examined to show this
in detail.

The construction requires a careful consideration
of the role of the identity operator in theories which
involve massive couplings. This can be done by introducing a coupling
associated with the identity operator, $\Lambda$
(a cosmological constant).
Its conjugate momentum,
${\phi_\Lambda}$, is just the expectation value
of the identity operator.
The $\beta$-function associated with $\Lambda$
turns out to be linear in $\Lambda$
\autoref\newcount\BC\BC=\refno,
$\beta^\Lambda(g,\Lambda)=-D\Lambda +U(g)$ where
$U(g)$ is an analytic function of the other couplings
and is independent of $\Lambda$.
The conventions adopted in this paper are such that
the cosmological constant is scaled
by its canonical dimensions and $\Lambda$ is dimensionless.
This explains the term $-D\Lambda$ here which not otherwise be
present. A consequence of this is that the $\phi_\Lambda=\kappa^D$
is also a density. For massless theories, $U(g)=0$.
For the purposes of the introductory discussion presented here
it will be assumed that the $\beta$-functions have no explicit
$\kappa$ dependence and only depend on $\kappa$ implicitly
through $g^{\hat a}(\kappa)$. A full generalisation to the
situation where a subtraction procedure is chosen in which
the $\beta$-functions have explicit $\kappa$ dependence
is given later.

It will be shown in section three that
the Hamiltonian defined by
$$H(g,\phi)=\beta^a(g)\phi_a +
\beta^\Lambda(g,\Lambda)\phi_\Lambda,\autoeq$$
\newcount\hbp\hbp=\count90
governs the renormalisation group flow
of the couplings $g^a$ and the expectation values $\phi_a$.
Of course, once
the theory has been solved, all of the vacuum expectation values
(VEV's) can be expressed as
functions of $g^a$ and $t$, $\phi_a=\phi_a(g,t)$ and then the
Hamiltonian can indeed then be written as a function of $g^a,\Lambda$
and $t$ alone (for example $\phi_\Lambda=\kappa^D$).
However the philosophy here is that
at the outset $g^a$ and $\phi_a$ are to
be considered as {\it independent} variables and
$H(g,\phi)$ depends on each separately.

The main result of this paper is that the RG evolution of $g^a$
and $\phi_a$ is given by \lq\lq Hamilton's equations",
$$\eqalign{{dg^a\over dt}&={\partial H\over\partial\phi_a}
\Vline\sub g\cr
{d\phi_a\over dt}&=-{\partial H\over\partial g^a}
\Vline\sub\phi.\cr}\autoeq$$
\newcount\hameq\hameq=\count90
The first equation follows simply from the definition of
$H(g,\phi)$ in (\the\hbp) while the second contains non-trivial
dynamics, despite the simple form of $H$. Indeed one can interpret
$U(g)$ as a potential and re-write
the second of equations (\the\hameq) as
$${d\phi\over dt}=-\kappa^DdU(g),\autoeq$$
\newcount\NII\NII=\count90
where $\phi=\phi_adg^a$ is a one-form and
$dU={\partial U\over\partial g^a}dg^a$ is the exterior derivative
of the potential.
The analogy between (\the\NII) and Newton's second law,
${dp\over dt}=-dU$ for a particle with momentum $p$ moving
in a potential $U$, is obvious.

The idea that RG flow might be related to a potential was first
suggested, to the author's knowledge, by Wallace and Zia
\autoref\newcount\WZia\WZia=\refno
but so far the investigations in this direction seem to have been
attempts to find a potential for the $\beta$-functions
which
requires introducing a metric on the space of couplings,
rather than for the VEV's as suggested here. In the construction
presented here a metric on $T({\cal M})$ is not necessary.

The reformulation of the RG evolution in terms of Hamiltonian flow
allows the introduction of Poisson brackets and their associated
symplectic
structure. (That a symplectic structure should be
relevant to RG flow was first suggested, to the authors knowledge,
in [\the\DenjoeChris].) Extending the set $\{g^a\}$ to include the
cosmological constant, $\{g^a,\Lambda\}=\{g^{\hat a}\}$ where
${\hat a}=1,\ldots,n$,
the Poisson bracket of any two quantities $A$ and $B$
is, of course,  given by expressing them as functions of
$g^{\hat a}$ and $\phi_{\hat a}$,
possibly also with an explicit $t$ dependence, and taking the
combination
$$\{A,B\}={\partial A\over\partial g^{\hat a}}
{\partial B\over\partial\phi_{\hat a}}
-{\partial A\over\partial\phi_{\hat a}}
{\partial B\over\partial g^{\hat a}}.\autoeq$$
Obviously
$$\{g^{\hat a},\phi_{\hat b}\}={\delta_{\hat a}}^{\hat b}.\autoeq$$
The RG evolution for any function on phase space is then given by
$${dA\over dt}={\partial A\over\partial t}\Vline\sub{g,\phi}
+\{A,H\}.\autoeq$$\newcount\rgpb\rgpb=\count90

Note in particular that, when there is no explicit $\kappa$ dependence
in the $\beta$-functions, the Hamiltonian (\the\hbp)
is a constant of the motion
${dH\over dt}=0$.
Once more we stress that,
even though it is trivial that ${\phi_\Lambda}=\kappa^D$,
it would be wrong to include
this $\kappa$ dependence in $H(g,\phi)$ explicitly -
$\phi_\Lambda$ is to be considered as an independent variable
in this formalism and $H$ has no explicit
$\kappa$ dependence.
One could, however, omit $\Lambda$ as an independent
variable and define a $t$-dependent Hamiltonian
$h(g,\phi,t)=\beta^a(g)\phi_a + \kappa^DU(g)$ on $T^*({\cal M})$.
This leads to the same equations of motion (\the\hameq)
but $h(g,\phi,t)$ is not a constant if there are massive
couplings in the theory, instead
${dh\over dt}={\partial h\over\partial t}=D\kappa^DU(g)$.

The classical analogy can be taken even further. It is also
argued in section three that the RG
equation for the generating function $w(g,t)$ can be written
as
$${\partial w\over\partial t}\Vline\sub g + H\left(g,{\partial
w\over\partial g}\right)=0,\autoeq$$
\newcount\HJ\HJ=\count90
which is clearly a field theoretic version of the Hamilton-Jacobi
equation.

The layout of the paper will be as follows. In \S 2 the RG equation
for $w(g,t)$ is derived, taking particular care over the role of the
identity operator and the way that the
cosmological constant is related to masses. In \S 3 a
Hamiltonian formalism of RG evolution is developed. A symplectic
structure on the phase space $(g^{\hat a},\phi_{\hat a})$ is introduced
and Hamilton's equations (\the\hameq)
are derived together with the Hamilton-Jacobi equation (\the\HJ).
The renormalisation group equation for $N$-point Green functions
is presented as a special case of equation (\the\rgpb).
\S 4 (which is the only part of this paper which uses perturbation
theory) exemplifies the ideas with the use of massive
$\lambda\varphi^4$ in four dimensions as a model. The implementation
of symmetries is discussed in \S 5, where some examples are used to
argue that Ward identities can be represented by symmetries on phase
space and can be used to construct RG invariants (\lq\lq constants
of motion") which commute with the Hamiltonian and generate the symmetry
through the Poisson bracket structure.
Finally the results are
summarised in \S 6.
\vfill\eject
{\bf \S 2 The Renormalisation Group Equation For The Partition
Function}
\vskip .5cm
Consider a renormalisable field theory, parameterised by a set of
renormalised couplings $g^a_R,\; a=1,\ldots,n-1$ where $n$ is finite
(the superscript $n$
will be reserved for a cosmological constant). In this section all
renormalised quantities will be denoted  by the
letter $R$ so as to try to keep the argument as clear as possible.
The bare couplings $g^a_0$ can be thought of as functions
$g^a_0(g_R,\epsilon)$ of the renormalised couplings plus
a regularisation parameter, $\epsilon$ (strictly speaking these
functions should also contain an explicit $\kappa$ dependence
since their total derivative with respect to $\kappa$ must vanish,
but this is not shown here). The bare couplings are analytic
functions of the
$g_R^a$ provided $\epsilon\ne 0$, but are singular in the
limit $\epsilon\rightarrow 0$. The transformation
$g_0^a\rightarrow g_R^a$ can be viewed as a co-ordinate
transformation on the $n-1$-dimensional space of theories, $\cal M$.
This co-ordinate transformation is singular in
the limit $\epsilon\rightarrow 0$ but, as long as the theory is
renormalisable, this is not a disaster and can be treated
consistently. For the moment the method of regularisation is left
open, one could for example use a cut-off $\Lambda_c$ and set
$\epsilon=\kappa/\Lambda_c$ or dimensional continuation with
$\epsilon=D-4$.

The action can be written as a linear combination of \lq\lq basic"
operators $\Phi^0_a$, which will include composite operators,
$$S_0(g_0,\Phi^0,\epsilon)=\int {\cal L}_0(g_0,\Phi^0(x))d^Dx
\qquad\hbox{where}\qquad{\cal L}_0=g^a_0\Phi^0_a(x).\autoeq$$
Strictly speaking, since space is taken to be Euclidean, this
is the energy rather than the action - but a Hamiltonian will
appear in the next section in a totally different context.
By abuse of language therefore $S_0$ will be called the action.

In massive $\lambda\varphi^4$ theory, for example, one would
have
$${\cal L}_0=k_0\partial_\mu\varphi_0\partial^\mu\varphi_0
+j_0\varphi_0+{1\over 2}m_0^2\varphi_0^2+
{\lambda_0\over 4!}\varphi_0^4
\autoeq$$
\newcount\LAG\LAG=\count90
with $\Phi^0_k=\partial_\mu\varphi_0\partial^\mu\varphi_0$,
$\Phi^0_j=\varphi_0$, $\Phi^0_{m^2}={1\over 2}\varphi_0^2$
and $\Phi^0_\lambda={1\over 4!}\varphi_0^4$ and four independent
couplings $k_0,j_0,m_0^2$ and $\lambda_0$. Without loss of
generality the fields can be re-scaled to set $k_0=1$. If necessary
the couplings can be made functions of position. Thus the
notation can be extended to allow source terms,
$j_0(x)\varphi_0(x)$, with $j_0(x)$ a function of position
which may be set to zero after all (functional) differentiations
have been carried out. More generally all of the couplings can be
made to depend on position so as to introduce sources for the
composite operators as well
\autoref\newcount\Vasiliev\Vasiliev=\refno.  
After all differentiations have been carried out these sources
can be set to constant values if so desired.

Returning to the general case, consider a renormalisable
theory written in terms of the bare couplings. Representing all of
the bare fields generically by $\varphi_0$, the partition function
(generating functional) is,
$$\Z_0(g_0)=\int{\cal D}\varphi_0\e^{-S_0(g_0,\Phi^0)}.\autoeq$$
The generating functional for connected Green functions
(the free energy) is
$W_0(g_0)=-\ln\Z_0(g_0)$.
As usual the free energy density is defined via
$W_0(g_0)=\int w_0(g_0,x)d^Dx$. The introduction of $w_0$ avoids
trivial volume divergences in the case of infinite space.
It has canonical mass dimension $D$.

For future convenience a coupling for the identity operator
will be included,
$$\Z(g_0,\Lambda_0)=\int{\cal D}\varphi_0\e^{-S_0(g_0,\Phi^0)
-\int d^Dx\Lambda_0}=\e^{-\int d^Dx\Lambda_0}\Z_0(g_0)\autoeq$$
\newcount\Zdef\Zdef=\count90
(there is no bare subscript on $Z(g_0,\Lambda_0)$ because, as
will be explained later, it is finite).
In analogy with general relativity
$\Lambda_0$ might be called a cosmological constant, but in
different physical situations it would have different physical
interpretations.

It is important to realise that $\Lambda_0$ is independent
of $g_0^a$ and plays the role of a new coupling for the identity
operator. Thus the set $\{g_0^a\}$ can be extended to
$\{g_0^{\hat a}\}=\{g_0^a,\Lambda_0\}$ where $\hat a=1,\ldots,n$ are
co-ordinates on a $n$-dimensional manifold $\widehat{\cal M}$.
$\Lambda_0$ can be included in the bare Lagrangian as
${\cal L}_0=g_0^a\Phi^0_a+\Lambda_0{\bf 1}$.
In terms of densities
$$W(g_0,\Lambda_0)=-\ln\Z(g_0,\Lambda_0)=\int w(g_0,\Lambda_0)d^Dx,
\autoeq$$
with $w(g_0,\Lambda_0)=w_0(g_0^a)+\Lambda_0$ linear in $\Lambda_0$.

When all couplings are independent of position and the theory is
translationally invariant $w(g_0,\Lambda_0)$ is independent of $x$ and
this will be assumed from now on.
In situations in which translational
invariance is not a symmetry of $D$-dimensional space,
one will need to introduce
extra terms involving the Riemann tensor into the action [\the\BC].

Expectation values of bare quantities can be obtained by
differentiating $w(g_0,\Lambda_0)$ with respect to the couplings,
$$<\Phi^0_a>=\Bigl<{\partial{\cal L}_0\over\partial g_0^a}\Bigr>
={\partial w\over\partial g_0^a}\qquad\hbox{and}\qquad
1=<{\bf 1}>={\partial w\over\partial\Lambda_0}.\autoeq$$
The bare operators
$\Phi^0_{\hat a}={\partial{\cal L}_0\over\partial g_0^{\hat a}}$ are
co-vectors on the space of couplings, i.e.
$\Phi:=\Phi^0_{\hat a}dg_0^{\hat a}$ is an operator valued one-form.
This notion of the basic operators being co-vectors on the
space of couplings is implicit in the work of Zamolodchikov
\autoref\newcount\Zam\Zam=\refno.
For a conformal field theory in two dimensions
$\Phi^0_{\hat a}$ would be the primary fields of the theory.
Expectation values, $dw=<\Phi^0_{\hat a}>dg_0^{\hat a}$,
are (exact) real valued one-forms on $\widehat{\cal M}$.

The bare operators can be written as linear combinations of
renormalised operators
$$\Phi^0_{\hat a}={Z_{\hat a}}^{\hat b}\Phi^R_{\hat b},\autoeq$$
where ${Z_{\hat a}}^{\hat b}$ is a matrix of renormalisation
constants (see for example
\autoref\newcount\ChengLi\ChengLi=\refno).
In general there will be operator mixing and
${Z_{\hat a}}^{\hat b}$
will not be diagonal.
If the bare action has no massive couplings it is not necessary to
include the identity operator in the list and a cosmological
constant can be omitted, since ${Z_a}^n=0\;(a\le n-1)$
when all couplings
are massless. But if there are massive couplings in the bare action then
some or all of the ${Z_a}^n$ will be non-zero and $\Lambda_0$
plays a crucial role.

Following up the idea that the transition from bare to renormalised
couplings can be implemented as a co-ordinate transformation on
$\widehat{\cal M}$, we
consider $g_0^{\hat a}(g_R,\epsilon)$ to be analytic functions of
$g_R^{\hat a}$ (keeping $\epsilon\ne 0$ for the moment).
Consider
$${\partial{\cal L}_0\over\partial g_R^{\hat a}}=
{\partial g_0^{\hat b} \over\partial g_R^{\hat a}}
{\partial{\cal L}_0\over\partial g_0^{\hat b}}=
{\partial g_0^{\hat b} \over\partial g_R^{\hat a}}
\Phi^0_{\hat b}.\autoeq$$
Clearly these should be related to the renormalised operators
$\Phi^R_{\hat a}$. In
fact
$${\partial g_0^{\hat b} \over\partial g_R^{\hat a}}
={(Z^{-1})_{\hat a}}^{\hat b}\autoeq$$
is the inverse of the operator mixing matrix
${Z_{\hat a}}^{\hat b}={\partial g_R^{\hat b} \over\partial
g_0^{\hat a}}$
\autoref\newcount\Tod\Tod=\refno. Thus
$${\partial{\cal L}_0\over\partial g_R^{\hat a}}
={(Z^{-1})_{\hat a}}^{\hat b}\Phi^0_{\hat b}=\Phi^R_{\hat a}.
\autoeq$$
The operator valued one-form $\Phi$ can now be expressed in
either co-ordinate system,
$$\Phi=\Phi^0_{\hat a}dg_0^{\hat a}=\Phi^R_{\hat a}dg_R^{\hat a},
\autoeq$$
and similarly the real valued one-form $dw$ is
$$dw=<\Phi>=<\Phi^0_{\hat a}>dg_0^{\hat a}=<\Phi^R_{\hat a}>dg_R^{\hat a}.
\autoeq$$
Thus
$${\partial w\over\partial g_R^{\hat a}}=
{\partial g_0^{\hat b} \over\partial g_R^{\hat a}}<\Phi^0_{\hat b}>=
{(Z^{-1})_{\hat a}}^{\hat b}<\Phi^0_{\hat b}>=<\Phi^R_{\hat a}>.
\autoeq$$\newcount\phidef\phidef=\count90
Partial derivatives here mean, of course, that all quantities are considered
to be functions of the renormalised couplings, and each
$g_R^{\hat a}$ is varied independently of the others.
When massive couplings are present $\Lambda_0$ must
be considered to be a function of the renormalised couplings.
This is necessary because the $\Phi_0^a$ then mix with the identity
operator under renormalisation.

Note that, since both $g_R^{\hat a}$ and $<\Phi^R_{\hat a}>$ in
equation (\the\phidef) are finite as the
regularisation parameter $\epsilon\rightarrow 0$, $w(g_R,t)$ considered
as a function of the renormalised couplings and the renormalisation
point $\kappa$ must also be finite as $\epsilon\rightarrow 0$.
This is why there is no bare subscript on the definition
of $\Z(g_0,\Lambda_0)$ in equation (\the\Zdef), [\the\BC], and
is one of the reasons for introducing a coupling for the identity
operator -
both $\Z_0(g_0^a)$ and $\Lambda_0$ seperately diverge as $\epsilon\rightarrow
0$
but the combination
$\Z(g_R^{\hat a})=\Z(g_0^{\hat a})=
\e^{-\int d^Dx\Lambda_0}\Z_0(g_0^a)$ is
finite. A finite, renormalised generating function,
$W_R(g_R^a,t)=\int w_R(g_R^a,t)d^Dx=-\ln\Z_R(g_R^a)$, can now
be defined by
$$\Z(g_0^{\hat a})=\e^{-\int d^Dx\Lambda_0}\Z_0(g_0^a)=
\Z(g_R^{\hat a},t)=\e^{-\kappa^D\int d^Dx\Lambda_R}\Z_R(g_R^a,t)\autoeq$$
(the factor $\kappa^D$ in the exponential is
in accord with the convention that
$\Lambda_R$ is
dimensionless).
Thus $w=w_0(g_0^a)+\Lambda_0=w_R(g_R^a,t)+\Lambda_R\kappa^D$
can now be interpreted as a finite function of the renormalised
couplings with the crucial property that
$${\partial w\over\partial g_R^a}=<\Phi_a^R>\qquad\hbox{and}\qquad
{\partial w\over\partial \Lambda_R}=<\Phi_n^R>
=<\kappa^D{\bf 1}>=\kappa^D.\autoeq$$

Note that $\Lambda_0$ is linear in $\Lambda_R$, a fact which
follows from
the observation that the identity operator ${\bf 1}=\Phi_n^0$
does not get renormalised. Thus
$${(Z^{-1})_{\hat a}}^{\hat b}=\left(
\matrix{ {(Z^{-1})_a}^b & {(Z^{-1})_a}^n \cr
              0         &         \kappa^D      \cr}\right)
\qquad\hbox{since}\qquad {\partial g_0^n\over\partial g_R^n}
={\partial \Lambda_0\over\partial \Lambda_R}=\kappa^D.\autoeq$$
Indeed even the $\beta$-function for $\Lambda_R$
only depends on $\Lambda_R$
linearly through canonical dimensions, since
writing
$$\Lambda_0=\kappa^D\bigl(\Lambda_R+F(g_R^a,\epsilon)\bigr),\autoeq$$
we have
$$\beta^\Lambda={d\Lambda_R\over dt}=-D\Lambda_R
+U(g_R^a)\autoeq$$
\newcount\bela\bela=\count90
where $U(g_R^a):=-{dF\over dt}$ depends only on the $g_R^a$ for
$a\le n-1$, not on $\Lambda_R$,
and is finite as $\epsilon\rightarrow 0$.
The fact that $U(g_R)$ is independent of $\Lambda$ can
be seen from the following argument. $U(g_R)$ is a quantum
correction to the canonical dimensions of the cosmological
constant and as such can be determined (in principle)
using perturbation theory and Feynman diagrams, but
$\Lambda$ cancels out of all Feynamn diagrams due
to the normalisation factor ${1\over\Z}$. Hence
$U(g_R)$ is independent of $\Lambda$.

To make contact with the familiar notions
of a perturbative analysis one can write
the bare couplings in terms of the renormalised couplings as
$$g_0^{\hat a}(g_R,\epsilon)=
g_R^{\hat a}+\Delta g^{\hat a}(g_R,\epsilon)\autoeq$$
where $\Delta g^{\hat a}$ is a correction (which diverges as
$\epsilon\rightarrow 0$), so that
$$\eqalign{
{\cal L}_0&=g_0^{\hat a}\Phi^0_{\hat a}=
(g_R^{\hat a}+\Delta g^{\hat a}){Z_{\hat a}}^{\hat b}
\Phi^R_{\hat b}\cr
&=g_R^{\hat a}\Phi^R_{\hat a}\quad +
\quad\hbox{counter terms}\quad(C.T.'s).\cr}
\autoeq$$
This gives
$$\e^{-W}=\Z(g_0,\Lambda_0)=\e^{-\int d^Dx\Lambda_0}
\int{\cal D}\varphi_0\e^{-\int d^Dxg_R^a\Phi^R_a(x)+C.T.'s},
\autoeq$$
where the identity operator has been included among the
counter terms. Absorbing a further term proportional to the
identity into the counter terms we have
$$\e^{-W}=\e^{-\kappa^D\int d^Dx\Lambda_R}
\int{\cal D}\varphi_0\e^{-\int d^Dxg_R^a\Phi^R_a(x)+C.T.'s}.
\autoeq$$
If desired, one can perform the functional integral over
renormalised fields rather than bare fields by setting
${\cal D}\varphi_0\rightarrow {\cal D}\varphi_R$, where
$\varphi_0=z^{1/2}\varphi_R$ with $z$ being the wave function
renormalisation factor, and then adding a further term
${1\over 2}\ln z\int d^Dx{\bf 1}$ to the counter terms.

Leaving perturbation theory behind and returning to the
general analysis, we are now in a position to write the
renormalisation group equation for the free energy,
$w(g_R^{\hat a},t)$. Since $w=w_0(g_0^a)+\Lambda_0$ and
all bare couplings are independent of the renormalisation point
$t=\ln\kappa$, we have
$${dw(g_R^a,\Lambda_R,t)\over dt}
=\beta^a{\partial w\over\partial g_R^a}
+\beta^\Lambda{\partial w\over\partial \Lambda_R}
+{\partial w\over\partial t}\Vline\sub{g_R,\Lambda_R}=0.
\autoeq$$
If we now
denote the VEV's by $\phi^R_a:=<\Phi^R_a>$,
this reads
(since ${\partial w\over\partial\Lambda_R}=\kappa^D=\phi^R_\Lambda$
and ${\partial w\over\partial g^a_R}
=\phi^R_a$)
$${\partial w(g_R^{\hat a},t)\over\partial t}\Vline\sub{g_R,\Lambda_R}
+\beta^a(g_R^a)\phi^R_a
+\beta^\Lambda(g_R^{\hat a})\phi_\Lambda=0.\autoeq$$
\newcount\erg\erg=\count90
This is the equation that will be used in the next section
to argue for Hamiltonian flow on $(g_R^{\hat a},\phi^R_{\hat a})$ space.

Alternatively, since $\Lambda_R$ only ever appears in this
equation linearly,
it can be eliminated
using equation (\the\bela) and $w=w_R(g_R^a)+\Lambda_R\kappa^D$,
with $w_R$ independent of $\Lambda_R$, to give
$${\partial w_R(g_R^a,t)\over\partial t}\Vline\sub{g_R}
+\beta^a(g_R^a)\phi^R_a +\kappa^DU(g_R^a)=0.\autoeq$$
Note that it is {\it not} true, in general, that
${dw_R(g^a_R,t)\over dt}=0$, since the presence of massive
couplings necessitates the introduction the function
$U(g)$ in a general renormalisation prescription.

This analysis has been a somewhat lengthy treatment of
concepts that are not new, but it has been included
in order to expose clearly the role of the cosmological
constant in theories with massive couplings as well as to set up the
notation.
\vfill\eject
{\bf \S3 Symplectic Structure And Hamiltonian Flow Of The RG
Equation}
\vskip .5cm
 In this section it will be shown that the RG equation derived for
the generating function in the previous section naturally admits a
symplectic structure with its concomitant Poisson
brackets, and the renormalisation group flow can be
obtained  from a Hamiltonian function on phase space in a manner
analogous to dynamical evolution in classical mechanics (but
with important differences).

The starting point is equation (\the\erg)
$$\phi_{\hat a}{dg^{\hat a}\over dt}
+ \left({\partial w\over\partial t}\right) = 0.
\autoeq$$\newcount\wRrg\wRrg=\count90
{}From now on the qualifier $R$ on $g_R^{\hat a}$ and $\phi^R_{\hat a}$ will be
omitted as all quantities will be renormalised, unless otherwise
indicated. The
generating function $w$ appearing in this section is always
$w=w_R+\kappa^D\Lambda_R$. It is stressed that $w$ is linear in
$\Lambda$.

To highlight the analogy with classical mechanics, we shall
define a function $H$,
$$H=-{\partial w\over\partial t},\autoeq$$
\newcount\Hdef\Hdef=\count90
so that (\the\wRrg) can be written
$$H(g,\phi)=\beta^a(g)\phi_a + \beta^\Lambda\phi_\Lambda.\autoeq$$
\newcount\otherHdef\otherHdef=\count90
(For the moment it will be assumed that the $\beta$-functions
have no explicit $\kappa$ dependence so that $H(g,\phi)$
has no explicit $t$ dependence - a generalisation
including explicitly $\kappa$ dependent $\beta$-functions is
given towards the end of this section.) The philosophy now is to
forget where the $\phi_{\hat a}$ came from and treat them as independent
variables. It is only after the theory has been solved
that we can use $\phi_{\hat a}=\partial_{\hat a}w$.

Consider the left hand side of equation (\the\wRrg) as a differential
$$\Theta=\phi_{\hat a}dg^{\hat a}-Hdt,\autoeq$$
\newcount\deftheta\deftheta=\count90
where $\Theta(g,\phi,t)$ is a one-form on the $2n+1$
dimensional space parameterised by $g^{\hat a},\phi_{\hat a}$ and $t$.
When the theory is solved, and $\phi_{\hat a}(g,t)$ is written
as an explicit function of $g^{\hat a}$ and $t$,
$\Theta=dw$ is exact, but when the $\phi_{\hat a}$
are treated as independent variables $\Theta$ is not exact.
The discussion now parallel's the treatment of classical
mechanics in
\autoref\newcount\Arnold\Arnold=\refno.
Just as the couplings $g^{\hat a}(t)$ evolve along the RG trajectories,
so do the expectation values $\phi_{\hat a}(t)$. Thus the RG trajectories
can be pictured as flow lines in $(g^{\hat a},\phi_{\hat a},t)$ space.
Treating $\phi_{\hat a}$ as independent variables construct the two-form
$$\eqalign{
\Omega&={1\over 2}\Omega_{ij}dx^i\wedge dx^j
 =d\Theta=d\phi_{\hat a}\wedge dg^{\hat a} - dH\wedge dt\cr
 &=d\phi_{\hat a}\wedge dg^{\hat a}-{\partial H\over\partial g^{\hat
a}}\Vline\sub\phi
dg^{\hat a}\wedge dt - {\partial H\over\partial \phi_{\hat a}}\Vline\sub g
d\phi_{\hat a}\wedge dt,\cr}\autoeq$$
where $\{x^i\}=\{g^{\hat a},\phi_{\hat a},t\}\;i=1,\ldots,2n+1$ are
co-ordinates
on the $2n+1$ dimensional space.
Now $\Omega$ can be written as an anti-symmetric matrix,
$$\Omega_{ij}=\left(\matrix{
0&-{\bf I}&-{\partial H\over\partial g^{\hat a}}\vline\sub\phi\cr
{\bf I}&0&-{\partial H\over\partial \phi_{\hat a}}\vline\sub g\cr
{\partial H\over\partial g^{\hat a}}\vline\sub\phi
&{\partial H\over\partial \phi_{\hat a}}\vline\sub g&0\cr}
\right),\autoeq$$
where ${\bf I}$ is the $n\times n$ identity matrix.
Since $\Omega_{ij}$ is an odd dimensional anti-symmetric matrix
it must have at least one zero eigenvalue
(it will be assumed that it has only
one, otherwise the restriction  of $\Omega$ to surfaces of constant
$t$ would result in a degenerate symplectic form on phase space).
The corresponding eigenvector, $\vec\xi$, is easily seen to be
$$\xi^i=\Bigl({\partial H\over\partial \phi_{\hat a}}\Vline\sub g,
-{\partial H\over\partial g^{\hat a}}\Vline\sub\phi,1\Bigr).\autoeq$$
It seems natural
to identify the flow lines of the vector field $\vec\xi$ with renormalisation
group trajectories, since $\Theta$ is exact when the theory is
solved and
$\phi_{\hat a}(g,t)$ are substituted into equation (\the\deftheta).
This requires
$${dg^{\hat a}\over dt}={\partial H\over\partial \phi_{\hat a}}\Vline\sub g
\qquad\hbox{and}\qquad
{d\phi_{\hat a}\over dt}=-{\partial H\over\partial g^{\hat a}}\Vline\sub\phi.
\autoeq$$
\newcount\RGhameq\RGhameq=\count90
Obviously ${\partial H\over\partial \phi_{\hat a}}\vline_g=\beta^{\hat a}$
by definition, so the first equation is certainly consistent.
For $\hat a=n$, the second equation reduces to the identity
${d\kappa^D\over dt}=D\kappa^D$.
The interpretation of the other equation hinges on the crucial
observation that the function $U(g)=\beta^\Lambda+D\Lambda$ is
an analytic function of the $g^a$ (independent of $\Lambda$).
Thus, from equation (\the\otherHdef)
$${d\phi_a\over dt}=-{\partial\over\partial g^a}\Bigl(\kappa^DU(g)\Bigr)
-\left({\partial\beta^b\over\partial g^a}\right)\phi_b\autoeq$$
or
$${d\phi_a\over dt}+(\partial_a\beta^b)\phi_b=-\kappa^D\partial_a U.\autoeq$$
\newcount\coriolis\coriolis=\count90
This is the renormalisation group equation for the RG evolution
of the vacuum expectation values of the basic operators of the
theory. The parallel with Newton's second law is obvious.
The matrix of anomalous dimensions $\partial_a\beta^b$ appears
as a pseudo-force (Coriolis force) and the function $U(g)$ is
a potential. For massless theories $U$ vanishes, thus massless
theories are analogous to free particle motion.
Just as in classical mechanics the Coriolis force can be eliminated
when the motion of the basis vectors is included in the equation.
Consider therefore the one-form $\phi=\phi_adg^a$. One has
$$\eqalign{
{d\phi\over dt}&=\left({d\phi_a\over dt}\right)dg^a + \phi_a{d(dg^a)\over
dt}\cr
&=\left({d\phi_a\over dt}+\phi_b{\partial\beta^b\over\partial g^a}
\right)dg^a,\cr}\autoeq$$
where we have used ${d(dg^a)\over dt}=d({dg^a\over
dt})=\partial_b\beta^adg^b$. Thus equation (\the\coriolis) can be written
in co-ordinate free notation as,
$${d\phi\over dt}=-\kappa^DdU(g),\eqno(\the\NII)$$
where $dU={\partial U\over\partial g^a}dg^a$ is the exterior
derivative.
Of course, once the theory is solved, the $\phi_a$ can be expressed
as explicit functions of $g^a$ and $t$, $\phi_a(g,t)$, so that
$${d\phi\over dt}=\left({\partial\phi_a\over\partial t}\Vline\sub g
+\beta^b\partial_b\phi_a + (\partial_a\beta^b)\phi_b\right)dg^a,
\autoeq$$
and equation (\the\coriolis) then becomes
$${\partial\phi_a\over\partial t}\Vline\sub g
+\beta^b\partial_b\phi_a + (\partial_a\beta^b)\phi_b
=-\kappa^D\partial_aU,\autoeq$$\newcount\CZ\CZ=\count90
which is a version of the RG equation for
the VEV's, including the anomalous dimensions and the
inhomogeneous term $-\kappa^D\partial_aU$ which arises due to masses.

Yet another way of expressing this is to observe that the
left hand side of (\the\CZ) involves the
Lie derivative, ${\cal L}_{\vec\beta}\phi$,
of the one-form $\phi_adg^a$ with respect to
the vector field $\vec\beta=\beta^a{\partial\over\partial g^a}$
\autoref\newcount\Lassig\Lassig=\refno.
Since $\phi$ is exact we have
$${\cal L}_{\vec\beta}\phi=d(i_{\vec\beta}\phi)
={\partial\over\partial g^a}\Bigl\{\beta^b(g)\phi_b(g,t)\Bigr\}dg^a,
\autoeq$$
where $i_{\vec\beta}\phi$ denotes the contraction of
$\vec\beta$ with the one-form $\phi$,
$i_{\vec{\beta}}\phi=\beta^a\phi_a$. Thus
another way of writing (\the\coriolis) is
$${\partial\phi\over\partial t}\Vline\sub g=-d\bigl\{\kappa^DU(g)+
\beta^b(g)\phi_b(g)\bigr\},\autoeq$$
where again $d=dg^a{\partial\over\partial g^a}$.

The analogy with classical mechanics can be taken further still.
The definition of $H$ in equation (\the\Hdef),
$$H(g,\phi) +{\partial w\over\partial t} =0,\autoeq$$
can be expressed as a partial differential
equation in the $n+1$ variables $(g^{\hat a},t)$.
Since, when the theory is solved,
$\phi_{\hat a}={\partial w\over\partial g^{\hat a}}$ we have
$${\partial w\over\partial t}\Vline\sub g +
H\Bigl(g,{\partial w\over\partial g}\Bigr)=0,\eqno(\the\HJ)$$
which is clearly an analogue of the Hamilton-Jacobi equation.
(That the RG equation ought to be expressable as a
Hamilton-Jacobi type equation was first suggested
to the author by Denjoe O'Connor
and Chris Stephens
\autoref\newcount\DCup\DCup=\refno.)
Thus the generating functional in quantum field theory (or free
energy density in statistical mechanics) is playing the role
of the action in classical mechanics (Hamilton's principal function).

This structure suggests a reformulation of the renormalisation group.
Instead of expressing the RG running in terms
of $\beta$-functions and couplings, which can be thought of as
co-ordinates on configuration space
(the tangent bundle $T(\widehat{\cal M})$) it may be useful to use instead
phase space variables (the co-tangent bundle $T^*(\widehat{\cal M})$).
Any quantity, $A$, should then be considered to be a
function of the $2n$ co-ordinates $(g^{\hat a},\phi_{\hat a})$ and possibly
also the renormalisation point $t=\ln\kappa$. The RG
evolution of $A(g,\phi,t)$ is then given by,
$${dA\over dt}={\partial A\over\partial g^{\hat a}}
            {\partial H\over\partial\phi_{\hat a}}-
{\partial A\over\partial\phi_{\hat a}}{\partial H\over\partial g^{\hat a}}
+{\partial A\over\partial t}\Vline\sub{g,\phi}
=\{A,H\}+{\partial A\over\partial t}\Vline\sub{g,\phi},
\autoeq$$
where $\{A,H\}$ is the usual Poisson bracket with
$\{g^{\hat a},\phi_{\hat b}\}={\delta^{\hat a}}_{\hat b}$.
Since there is no explicit $\kappa$ dependence in
the Hamiltonian (\the\otherHdef) $H(g,\phi)$ is a RG invariant
(a constant of the motion)
$${dH\over dt}=0,\autoeq$$
but this is only true when there is no explicit $\kappa$ dependence
in the $\beta$-functions.

In particular the RG evolution of $N$-point Green functions
is of key importance in any theory. These can be viewed as
rank $N$ tensors on the space of couplings,
$$
G^{(N)}_{a_1\cdots a_N}(x_1,\ldots,x_N)=
<\tilde\Phi_{a_1}(x_1)\cdots\tilde\Phi_{a_N}(x_N)>,
\autoeq$$
where
$\tilde\Phi_{a_i}(x_i)=\Phi_{a_i}(x_i)-\phi_{a_i}$ has zero
vacuum expectation value (and is independent of $\Lambda$). The RG equation for
$G^{(N)}_{a_1\cdots a_N}(x_1,\ldots,x_N)$ is obtained by the process
described above,
$$\eqalign{
{d\over dt}&G^{(N)}_{a_1\cdots a_N}(x_1,\ldots,x_N)=\cr
&\left({\partial\over\partial t}\Vline\sub{g,\phi}
+\beta^b{\partial\over\partial g^b}\Vline\sub{\phi,t}
-\phi_c\left({\partial\beta^c\over\partial g^b}\right)
{\partial\over\partial\phi_b}\Vline\sub{g,t}
-\kappa^D\partial_b U{\partial\over\partial\phi_b}\Vline\sub{g,t}
\right)
G^{(N)}_{a_1\cdots a_N}(x_1,\ldots,x_N).\cr}\autoeq$$
If the co-vector basis $dg^a$ is also included, so as to write
the tensor in co-ordinate free notation
\hbox{$G^{(N)}=G^{(N)}_{a_1\cdots a_N}(x_1,\ldots,x_N)
dg^{a_1}\cdots dg^{a_N}$},
one arrives at the equation
$$\eqalign{
\left({d G^{(N)}\over dt}\right)_{a_1\cdots a_N}
={\partial\over\partial t}G^{(N)}_{a_1\cdots a_N}
&+\beta^b{\partial\over\partial g^b}\Vline\sub{\phi,t}G^{(N)}_{a_1\cdots a_N}
+\sum_{i=1}^N(\partial_{a_i}\beta^b)
G^{(N)}_{a_1\cdots a_{i-1}ba_{i+1}\cdots a_N}\cr
&-\phi_c(\partial_b\beta^c)
{\partial\over\partial\phi_b}\Vline\sub{g,t}G^{(N)}_{a_1\cdots a_N}
-\kappa^D\partial_b U{\partial\over\partial\phi_b}\Vline\sub{g,t}
G^{(N)}_{a_1\cdots a_N}.\cr}\autoeq$$
\newcount\pRGN\pRGN=\count90
The RG equation for $N$-point Green functions in the Hamiltonian
formalism is finally obtained
by observing that $G^{(N)}$ can equally
well be written in bare co-ordinates and so is independent of $t$.
The left hand side of (\the\pRGN) therefore vanishes,
${dG^{(N)}\over dt}=0$, and the RG equation is
$$\eqalign{
{\partial\over\partial t}G^{(N)}_{a_1\cdots a_N}(g,\phi,t)&
+\beta^b\partial_bG^{(N)}_{a_1\cdots a_N}(g,\phi,t)
+\sum_{i=1}^N(\partial_{a_i}\beta^b)
G^{(N)}_{a_1\cdots a_{i-1}ba_{i+1}\cdots a_N}(g,\phi,t)\cr
&=\Bigl(\phi_c(\partial_b\beta^c)
+\kappa^D\partial_b U\Bigr){\partial\over\partial\phi_b}
G^{(N)}_{a_1\cdots a_N}(g,\phi,t).\cr}\autoeq$$
\newcount\RGN\RGN=\count90
Alternatively, using equation (\the\CZ)
and allowing for the $g^a$ and $t$ dependence of $\phi_a$
after the theory has been solved,
this equation can be re-expressed as
$$\eqalign{
{\partial\over\partial t}&G^{(N)}_{a_1\cdots a_N}(g,\phi,t)
+\beta^b{\partial\over\partial g^b}
G^{(N)}_{a_1\cdots a_N}(g,\phi,t)
+\beta^c\left({\partial\phi_b(g,t)\over\partial g^c}\right)
{\partial\over\partial\phi_b}
G^{(N)}_{a_1\cdots a_N}(g,\phi,t)\cr
&+\sum_{i=1}^N(\partial_{a_i}\beta^b)
G^{(N)}_{a_1\cdots a_{i-1}ba_{i+1}\cdots a_N}(g,\phi,t)
+\left({\partial\phi_b(g,t)\over\partial t}\right)
{\partial\over\partial\phi_b}G^{(N)}_{a_1\cdots a_N}(g,\phi,t)=0
.\cr}\autoeq$$
\newcount\NRG\NRG=\count90
Were it not for the last term on the left hand side of this
equation, it would just be the definition of the Lie
derivative of $G^{(N)}$ with respect to the vector field
$\vec{\beta}$ - the last term is a correction to this
interpretation.
That the RG equation for $N$-point amplitudes
could be written as a Lie derivative was observed in
[\the\Lassig], and
corrections to this interpretation were investigated in
\autoref\newcount\HughIan\HughIan=\refno.

Returning now to the general formalism let us consider
more general canonical transformations.
A renormalisation group transformation is like time evolution
in classical mechanics and as such preserves the symplectic
structure on the $2n$-dimensional space
$(g^{\hat a},\phi_{\hat a})$,
$$\widehat\omega=\Omega\vert_{t=\hbox{const}} = d\phi_{\hat a}\wedge dg^{\hat
a}.
\autoeq$$
\newcount\symp\symp=\count90
Of course, there will in general be other important canonical
transformations which do not necessarily correspond to RG
transformations. In particular, the transformation from bare to
renormalised couplings preserves $\widehat\omega$,
$$\widehat\omega= d\phi_{\hat a}\wedge dg^{\hat a}
= d\phi^0_{\hat a}\wedge dg_0^{\hat a},\autoeq$$
but in a general renormalisation prescription
this would not correspond to a RG transformation
(except perhaps if the theory is regularised by using a cut-off
and BPHZ renormalisation, in which case the bare couplings do
have the interpretation of just being the renormalised
couplings at some very high energy, $\Lambda_c$).
Since the bare couplings are independent of the renormalisation
point, ${dg_0^{\hat a}\over dt}={d\phi^0_{\hat a}\over dt}=0$,
this transformation
is analogous to the canonical transformation in classical mechanics
which takes one from time dependent phase space variables
$(q^a(t),p_a(t))$ to the initial point $(q_0^a,p^0_a)$.
Referring back to the Hamilton-Jacobi equation (\the\HJ)
we see that the generating function for this canonical
transformation, in the familiar classical mechanical sense,
is nothing other than the generating function(al)
of the quantum field theory, $w$.

The analogy with the Hamilton-Jacobi equation of classical
mechanics can be further highlighted by explicitly indicating that
$w(g,t)$ depends on the subtraction procedure and writing it
as $w(g(t),g_0,t)$. This emphasises its dependence on the
counterterms $\Delta g^{\hat a}=g_0^{\hat a}-g^{\hat a}(t)$,
and does not affect the argument that ${dw\over dt}=0$.
In BPHZ renormalisation $g_0^{\hat a}$ really can be thought
as lying on on the RG trajectory and the analogy
between $w(g(t),g_0,t)$ and Hamilton's principal function,
the action $S(q,q_0,t)$ along a classical trajectory is even stronger.

A crucial difference between the phase space approach to the
RG presented here and classical mechanics
lies in
the Legendre transform,
$$H(g,\phi)-\beta^{\hat a}\phi_{\hat a}=0,\autoeq$$
\newcount\LT\LT=\count90
with $\beta^{\hat a}={\partial H\over\partial\phi_{\hat a}}\vert_g$,
which vanishes and in particular is not invertible.
However, as the examples of the next two sections
show, the flow is still far from trivial!

The Legendre transform presented here is also much simpler in form than
quantum field theory Legendre transform introduced by Jona-Lasinio
\autoref\newcount\JL\JL=\refno.\footnote*{The
usual Legendre transform involves only one operator
$j(x)\varphi(x)$, composite operators being obtained by multiple
functional differentiation at the same point. But this can be
extended to include sources for composite operators
[\the\Vasiliev].}
The latter is implemented at the level of the generating functional
$w$ itself, rather than on its derivative
${\partial w\over\partial t}$, and this leads to the effective
action, which most certainly does not vanish.

Note that the Hamiltonian (\the\otherHdef),
$$H=\beta^a\phi_a + \beta^\Lambda\phi_\Lambda,\autoeq$$
actually has a simple physical interpretation.
The right hand side of this equation is just the negative of the usual
definition of the vacuum expectation value of the trace of
the energy-momentum tensor of the theory, $H=-<T>$.
It should not come as a surprise that
$<T>={\partial w\over\partial t}\vline_g$ since varying $t$
with the couplings fixed is completely equivalent
to a conformal rescaling of the metric.
The derivative ${\partial\over\partial t}\vline_g$ acting on
$w$ simply pulls down the action from the exponent and then
varies the metric leading to $<{T^\mu}_\mu>$.
Thus {\it the
entire RG evolution is governed solely by} $<T>$.

At fixed points of the RG flow (conformal field theories)
the Hamiltonian vanishes, because the $\beta$-functions do.
However once the theory is solved and explicit
expressions for $\phi_{a}$
in terms of $g^a(t)$ and $t$ are substituted into the
Hamiltonian, the resulting function is not analytic
at fixed points.
Derivatives higher than the first may be singular, as the example
of $\lambda\varphi^4$ treated in the next section
shows. This is to be expected since the Hamiltonain
is defined in terms of the free energy density which
is non-analytic at critical points.

The analysis so far has assumed that a subtraction
procedure is chosen so that the $\beta$-functions
only depend on $\kappa$ implicitly through $g^{\hat a}(\kappa)$
and have no explicit $\kappa$ dependence. Sometimes, however,
it may be convenient to use a subtraction procedure which results
in $\beta$-functions which have an explicit $\kappa$ dependence,
$\beta^{\hat a}(g,t)$. This can be incorporated into the
present framework by considering $t$ to be like an
extra coupling and extending the $n$-dimensional manifold
$\widehat{\cal M}$ to a $n+1$-dimensional manifold
$\widehat{\cal M}_E$ with $t$ as the extra co-ordinate.
The momentum conjugate to $t$ is (minus) the Hamiltonian,
$\phi_t=\partial_tw=-H(g,\phi,t)$. The $2n$-dimensional phase space
$T^*(\widehat{\cal M})$ is now extended to a $2n+2$-dimensional
phase space $T^*(\widehat{\cal M}_E)$. By definition one has $\beta^t=1$.
This is clearly analogous to the
situation in classical mechanics where phase space is
extended to include the energy and time as extra co-ordinates.
The Hamiltonian on the extended phase space is
$${\cal H}_E(g,\phi,t)=\beta^{\hat a}(g,t)\phi_{\hat a}
+\phi_t,\autoeq$$
and a new evolution parameter $\tau$ is introduced which is
ultimately identified with $t$ when the theory is solved.
When the theory is solved one has ${\cal H}_E=0$, which
is just the Hamilton-Jacobi equation for
$H(g,\phi,t)=\beta^{\hat a}\phi_{\hat a}$,
$${\partial w\over\partial t}+
H\Bigl(g,{\partial w\over\partial g},t\Bigr)=0.\autoeq$$
\newcount\HamJac\HamJac=\refno
On the extended phase space Hamilton's equations are
supplemented by
$${d\phi_t\over d\tau}=
-{\partial {\cal H}_E\over\partial t}\Vline\sub{g^{\hat a},\phi}
=-(\partial_t\beta^{\hat a})\phi_{\hat a}.\autoeq$$
In other words, with $\tau=t$,
${dH\over dt}=
{\partial H\over\partial t}=\partial_t\beta^{\hat a}\phi_{\hat a}$
and the $t$ dependent Hamiltonian $H(g,\phi,t)$ on
$T^*(\widehat{\cal M})$ is not a RG invariant when such
subtraction procedures are used. Apart from this difference
the analysis is the same as before and the evolution can
be described on $T^*(\widehat{\cal M})$ with explicitly
$t$ dependent $\beta$-functions.

Finally one might ask, what is the special ingredient of
renormalisation group flow which allows it to be written
in Hamiltonian form? After all, one cannot expect any
vector flow to be expressible as a Hamiltonain flow.
The crucial ingredient is the fact that ${dw\over dt}=0$.
Thus the renormalisation group is, as the name implies, a symmetry.
This leads to the existence of the one-form $\Theta$, which
reflects this symmetry, as it is constant along
the RG flow, ${\cal L}_{\vec{\xi}}\Theta=0$ since
$i_{\vec{\xi}}\Theta=0$.
In particular, as stated at the beginning of this section,
${dw_R\over dt}\ne 0$ and one cannot, in general, get
Hamiltonian flow if $\Lambda$ is ignored.
Note that restricting $\Theta$ to surfaces of constant
$t$, $\widehat\theta=\Theta_{t=\hbox{const}}$ gives
$i_{\vec\xi}\widehat\theta=H$,
as expected for a Hamiltonian vector field.

\vfill\eject
{\bf \S 4 An Example - Massive $\lambda\varphi^4$}
\vskip .5cm
The construction of the previous section will now be applied to the
example of massive $\lambda\varphi^4$.
We shall use dimensional regularisation and
minimal subraction in $D=4-\epsilon$ dimensional Euclidean space.
We start with the Lagrangian (\the\LAG),
with a coupling for the identity operator included and $k_0$ set to unity,
$$ {\cal L}_0={1\over 2}\partial_\mu\varphi_0\partial^\mu\varphi_0
+j_0\varphi_0+{1\over 2}m_0^2\varphi_0^2+
16\pi^2{\lambda_0\over 4!}\varphi^4+\Lambda_0{\bf 1}\autoeq$$
\newcount\philag\philag=\count90
(the factor $16\pi^2$ is inserted to tidy up some later formulae -
it is $\lambda_0$ which is really the expansion parameter
in $\varphi^4$ theory).
It will be assumed that $m_0^2>0$ and that radiative corrections
do not change this.
A cubic term has been omitted because it can always be eliminated
by shifting the field $\varphi_0\rightarrow\varphi+const.$
(independently of the scaling that sets $k_0=1$).

The scalar field $\varphi_0$ has canonical dimension
$1-{\epsilon\over 2}$
and the couplings $j_0,m_0^2,\lambda_0$ and $\Lambda_0$
have canonical (mass) dimensions $3-{\epsilon\over 2},2,\epsilon$
and $4-\epsilon$ respectively.
For simplicitly we
shall restrict ourselves to the consideration of theories
symmetric in $\varphi_0\rightarrow-\varphi_0$ and set $j_0=0$.

The regularisation of the composite operators follows the
analysis of [\the\BC],
but note that here $\Lambda_0$ has the opposite sign to that
reference. On dimensional grounds one expects
$$\Lambda_0=\kappa^{4-\epsilon}\bigl(\Lambda
+ m^4F(\lambda,\epsilon)\bigr),\autoeq$$
for some function $F(\lambda,\epsilon)$, analytic in the renormalised
coupling $\lambda$ when $\epsilon\ne 0$ (as stated earlier all the
renormalised couplings are rescaled to be dimensionless by
multiplying by canonical powers of $\kappa$ - thus
$m^2=\kappa^{-2}\tilde{m}^2$ where $\tilde{m}^2$ is the usual
mass with canonical dimension two).

In dimensional regularisation the renormalised couplings are
given in terms of the bare couplings by
$$m^2=\kappa^{-2}z_2(\lambda,\epsilon)m_0^2\qquad
\lambda=\kappa^{-\epsilon}z_\lambda(\lambda,\epsilon)\lambda_0$$
$$\Lambda=\kappa^{-4+\epsilon}\Lambda_0 -
\kappa^{-4}m_0^2z_2^2(\lambda,\epsilon)F(\lambda,\epsilon),\autoeq$$
\newcount\cotran\cotran=\count90
where $z_2$ and $z_\lambda$ are the usual renormalisation
co-efficients for the mass and $\varphi^4$ coupling.
Had the linear source been retained, wave function renormalisation
would also have entered via $j=\kappa^{-3+\epsilon/2}z_1j_0$.

The $\beta$-functions are immediate,
$$\eqalign{
\beta^\lambda =\kappa{d\lambda\over d\kappa}=-\epsilon
+\beta(\lambda)
&\qquad\hbox{with}\qquad{\beta(\lambda)\over\lambda}=z_\lambda^{-1}
\kappa{dz_\lambda\over d\kappa}\cr
\beta^{m^2}=
\kappa{dm^2\over d\kappa}=\bigl(-2+\delta(\lambda)\bigr)m^2
&\qquad\hbox{with}\qquad\delta(\lambda)=z_2^{-1}\kappa{dz_2\over
d\kappa}\cr
\beta^\Lambda=\kappa{d\Lambda\over d\kappa}=(-4+\epsilon)\Lambda
+m^4\zeta(\lambda)
&\qquad\hbox{with}\qquad\zeta(\lambda)=
\bigl(\epsilon-2\delta(\lambda)\bigr)F-\kappa{dF\over d\kappa}.\cr}
\autoeq$$
\newcount\betafunctions\betafunctions=\count90
In particular the potential factorises in minimal subtraction,
$U(\lambda,m^2)=m^4\zeta(\lambda)$.
These $\beta$-functions
can be supplemented by wave function renormalisation
which may be obtained from the $\beta$-function for $j$,
$$\beta^j=\kappa{dj\over d\kappa}=\Bigl(-3+{\epsilon\over
2}+\gamma(\lambda)\Bigr)
\qquad\hbox{with}\qquad\gamma(\lambda)=z_1^{-1}\kappa{dz_1\over
d\kappa}\autoeq$$
and $j$ is set to zero after $\gamma(\lambda)$ has been
extracted.
Of course $\epsilon$ can be put to zero in all of these
$\beta$-functions since everything is finite. As is well known,
the functions can be developed in a power series in
$\lambda$
$$\eqalign{
\beta(\lambda)&=3\lambda^2 + o(\lambda^3)\cr
\gamma(\lambda)&={1\over 6}\lambda^2 + o(\lambda^3)\cr
\delta(\lambda)&=\lambda-{5\over 6}\lambda^2 + o(\lambda^3)\cr
\zeta(\lambda)&={1\over 32\pi^2}\Bigl(1+{\lambda^2\over 8} +
o(\lambda^3)\Bigr)\cr}\autoeq$$
\newcount\anb\anb=\count90
(for the $\lambda^2$ term in $\zeta(\lambda)$
see
\autoref\newcount\Hathrell\Hathrell=\refno.)
{}From equations (\the\cotran) one has
$${(Z^{-1})_{\hat a}}^{\hat b}=
{\partial g_0^{\hat b}\over \partial g^{\hat a}}
=\left(
\matrix{
-{\epsilon\lambda
z_\lambda^{-1}\over(-\epsilon\lambda + \beta)}\kappa^\epsilon
  &-{m^2\delta z_2^{-1}\over(-\epsilon\lambda + \beta)}\kappa^2
     &{m^4\{ F(\epsilon -2\delta)+\zeta \}\over
      (-\epsilon\lambda + \beta)}\kappa^{4-\epsilon}\cr
0  &  z_2^{-1}\kappa^2  &  2m^2F\kappa^{4-\epsilon}  \cr
0  &  0  &  \kappa^{4-\epsilon}  \cr}
\right).\autoeq$$
Thus the renormalised operators are
$$\eqalign{
&\Phi_\lambda=-\left({\epsilon\lambda
             z_\lambda^{-1}\kappa^\epsilon\over-\epsilon\lambda
             + \beta}\right)
             {16\pi^2\over 4!}\varphi^4_0
             -\left({m^2\delta z_2^{-1}\kappa^2\over-\epsilon\lambda
             + \beta}\right)
             {\varphi^2_0\over 2}
             +\left({m^4\{F(\epsilon -2\delta)+\zeta\}\kappa^{4-\epsilon}\over
             -\epsilon\lambda + \beta}\right)\;{\bf 1} \cr
&\Phi_{m^2}=z_2^{-1}\kappa^2{\varphi^2_0\over 2} +
        2m^2F\kappa^{4-\epsilon}{\bf 1}\cr
&\Phi_\Lambda=\kappa^{4-\epsilon}{\bf 1}\cr
}\autoeq$$
\newcount\regphi\regphi=\count90
(strictly speaking the first of these equations should
have a term on the right hand side involving the equations
of motion for the field [\the\BC], but since these operators will only
be used inside expectation values, this is omitted).

In order to evaluate expectation values of the renormalised
operators one has to interpret the meaning of the
operators $\varphi_0^4$ and $\varphi^2_0$. In a functional
integral approach, these would be the objects appearing in the
exponential. When brought down by differentiation they should
be considered as time ordered products in expectation values,
but of course there are divergences because they consist
of products of the field at the same point. These divergences
can be treated in dimensional regularisation in the usual
way, by using Wick's theorem for free fields and expanding
expectation values perturbatively. For example
for the operator $\varphi_0^2(x)$ Wick's theorem gives
$${\varphi_0^2(x)\over 2}={:\varphi_0^2(x):\over 2}
+{1\over 2}D(0),\autoeq$$
where $:\varphi_0^2(x):$ denotes normal ordering (with respect
to the renormalised mass) and
$$D(x)=<\varphi_0(x)\varphi_0(0)>=\int{d^Dp\over(2\pi)^D}
{\e^{-ip.x}\over(p^2 +m^2)}.\autoeq$$
Performing the integral using dimensional continuation yields
$$D(0)={1\over 4\pi}\left({m^2\over 4\pi}\right)^{1-{\epsilon\over 2}}
\Gamma\Bigr(-1+{\epsilon\over 2}\Bigr).\autoeq$$
Thus the expectation value is
$${1\over 2}<\varphi_0^2>=
{1\over 8\pi}\left({m^2\over 4\pi}\right)^{1-{\epsilon\over 2}}
\Gamma\Bigr(-1+{\epsilon\over 2}\Bigr)
+ o(\lambda),\autoeq$$
and the higher order terms can be obtained by expanding the
exponential in the functional integral and performing the
space integrals (note that $<:\varphi_0^2(x):>=o(\lambda)$ is
not zero at higher orders because the normal ordering is only defined
for free fields).
Including the identity term (\the\regphi) in the expression for
$\Phi_{m^2}$ one finds that a pole in $F$ cancels the pole
in $D(0)$ (as it must do) and, after setting $\epsilon$ to zero
$$<\Phi_{m^2}>={\kappa^4m^2\over 32\pi^2}(-1+\ln m^2)+o(\lambda),
\autoeq$$
where the $\overline{MS}$ scheme has been used so as to
avoid $-\gamma +\ln4\pi$ terms.

Including the next order contribution gives (after some work)
$$\eqalign{
&\phi_\lambda=<\Phi_\lambda>=
{\kappa^4m^4\over 128\pi^2}(1-\ln m^2)^2 + o(\lambda)\cr
&\phi_{m^2}=<\Phi_{m^2}>
={\kappa^4m^2\over 32\pi^2}
\Bigl\{-1+\ln m^2 + {\lambda\over 2}
\bigl((\ln m^2)^2 - \ln m^2\bigr)+o(\lambda^2)\Bigr\}\cr
&\phi_\Lambda=<\Phi_\Lambda>=\kappa^4\qquad
\hbox{to all orders}.\cr}\autoeq$$
\newcount\VEVs\VEVs=\count90
These can be derived from the generating functional
$$w(\lambda,m^2,\Lambda,t)={m^4\kappa^4\over 64\pi^2}
\Bigl(-{3\over 2}+\ln m^2 + {\lambda\over 2}(1-\ln m^2)^2
+ o(\lambda^2)\Bigr)+\kappa^4\Lambda\autoeq$$
\newcount\genfunct\genfunct=\count90
by $\phi_{\hat a}={\partial w\over\partial g^{\hat a}}$.
The calculation of the order $\lambda^2$ terms in $w$
would require extracting the finite part of the three loop
diagram $\eye$.

The Hamiltonian is thus
$$\eqalign{
H(g,\phi)&=\beta^a(g)\phi_a + \beta^\Lambda\phi_\Lambda\cr
&=\beta(\lambda)\phi_\lambda +
\bigl(-2+\delta(\lambda)\bigr)m^2\phi_{m^2}+
\bigl(m^4\zeta(\lambda)-4\Lambda\bigr)\phi_\Lambda.\cr}\autoeq$$
It is not difficult to prove, from
equations (\the\betafunctions), (\the\anb) and (\the\VEVs),
that this is a renormalisation group invariant,
${dH\over dt}=0$, to order $\lambda$.
In fact one finds
$H=-{\partial w\over\partial t}=-4w$
when the explicit solutions for the VEV's (\the\VEVs) are
substituted into the Hamiltonian. This is a consequence of
the fact that all couplings have been scaled to be dimensionless
so that the only explicit $\kappa$ dependence in
(\the\genfunct) is the trivial volume factor.

The renormalisation group evolution of the vacuum expectation
values is thus given by
$$\eqalign{
{d\phi_{m^2}\over dt}&+(\partial_{m^2}\beta^a)\phi_a =
-\kappa^D\partial_{m^2}U\cr
{d\phi_\lambda\over dt}&+(\partial_\lambda\beta^a)\phi_a =
-\kappa^D\partial_\lambda U,\cr}\autoeq$$
where the potential is
$U(g)={m^4\over 32\pi^2}+o(\lambda^2)$.
Using the explicit expressions for $\phi_a$ in (\the\VEVs) and the
$\beta$-functions, (\the\betafunctions) and (\the\anb),
it is a straightforward
calculation to check these equations explicitly to order $\lambda$.

Three comments on the analysis presented here are in order.

\noindent (i) Using dimensionless couplings will always give
${\partial w\over\partial t}\vline_g=4w$ Thus
$H=-4w$. However, it must be
stressed that in applying Hamilton's equations $g^{\hat a}$ and
$\phi_{\hat a}$
in $H(g,\phi)$ must be considered as {\it independent} variables.
It is only after the theory has been solved and $\phi_{\hat a}(g,t)$
determined as a function of $g^{\hat a}$ and $t$ that
$H$ can be identified with $-4w$.
\smallskip
\noindent (ii) The Hamiltonian,
$H=-<T>$, is a constant of the motion
(RG invariant).
This is consisted with the observation that
conservation of momentum implies that the energy-momentum
tensor does not get renormalised (provided the subtraction procedure
is compatable with translational
invariance
\autoref\newcount\CW\CW=\refno,
as $\overline{MS}$ is). Thus the bare energy-momentum operator
is equal to the renormalised one and $<T_0>=<T>$.
(The energy-momentum operator here is, of course, the \lq\lq improved"
operator of reference [\the\CW], obtained by coupling the
the scalar field $\varphi_0$ to the curvature scalar $\cal R$
of $D$-dimensional space before varying the metric and only
setting ${\cal R}=0$ afterwards.) How then can the statement
that the bare couplings are RG invariants be reconciled
with the fact that $-H=<T>=<T_0>\ne 0$? One answer is that
$H$ has canonical
mass dimension $D$ despite being a RG invariant.
The dimensionless
Hamiltonian $\tilde H=\kappa^{-D}H$
vanishes at high energy. Perhaps a more rigorous way
of stating this is to observe that, as we are using
dimensionless renormalised couplings so we could
also use dimensionless bare couplings
$$\tilde\Lambda_0=\kappa^{4-\epsilon}\Lambda_0,
\qquad \tilde m_0^2=\kappa^2m_0^2,
\qquad\tilde\lambda_0=\kappa^\epsilon\lambda_0,\autoeq$$
where the quantities with the tildes are the dimensionful ones.
One then obtains
$${dw(g_0,\phi_0)\over dt}=\beta_0^{\hat a}\phi^0_{\hat a}+<T_0>=0.
\autoeq$$
Since $\beta_0^{\hat a}$ are now non-zero (they are
simply the negative of the canonical dimensions of the couplings)
this equation is quite consistent. For a regularisation procedure
which does not preserve translational invariance
one looses the Ward identity that protects the energy-momentum
operator from renormalisation and it is no longer
necessary to have $<T_0>=<T>$.
\smallskip
\noindent (iii) Setting $k_0=1$ and then forgetting about it may
at first sight seem a little dangerous. After all it is clearly
related to wave-function renormalisation and should run like
all the other couplings. The fact that the operator
$\partial^\mu\varphi_0\partial_\mu\varphi_0$ can be ignored
with impunity is related to the equations of motion
(the Schwinger-Dyson equation). There is always one
linear combination of the basic operators $\Phi^0_a$ which
does not get renormalised, namely that corresponding to the
\lq\lq equations of motion"
$E_0(x)=\varphi_0(x){\delta S_0\over\delta\varphi_0(x)}$,
$$E_0=-\varphi_0\Box\varphi_0 + m_0^2\varphi_0^2 +
(16\pi^2){\lambda_0\over 3!}\varphi_0^4.\autoeq$$
There is a \lq\lq Ward identity" (the Schwinger-Dyson equation)
which ensures that this combination of linear operators
does not get renormalised, so $E_R=E_0$. For this reason one
of the operators in the original Lagrangian is always
redundant and can be ignored, and
here it is
$\partial^\mu\varphi_0\partial_\mu\varphi_0$ that has been ignored.
Strictly speaking though
$\partial^\mu\varphi_0\partial_\mu\varphi_0$
is not the same operator as
$-\varphi_0\Box\varphi_0$ and they should be treated separately.
A complete analysis is given in [\the\BC],
but these complications are omitted here in the
interests of clarity.
\vfill\eject
{\bf \S 5 Symmetries}\hfill
\vskip .5cm
The  notion of a Poisson bracket structure for the renormalisation group
evolution, as introduced in section three, immediately raises the question of
how symmetries might be implemented on the phase space
$(g^{\hat a},\phi_{\hat a})$.
As a first, almost trivial, example of a symmetry consider a
$N$-component scalar field $\varphi_0^i,\;i=1,\ldots,N$ with
Lagrangian
$${\cal L}_0={1\over 2}\sum_{i=1}^N
\partial_\mu\varphi_0^i\partial^\mu\varphi_0^i
+{1\over 2}m_0^2\sum_{i=1}^N(\varphi_0^i)^2
+{1\over 4!}\sum_{ijkl}\lambda_0^{ijkl}\varphi_0^i
\varphi_0^j\varphi_0^k\varphi_0^l.\autoeq$$
There are in general ${1\over 4!}N(N+1)(N+2)(N+3)$ different
couplings $\lambda_0^{ijkl}$ and each of these could renormalise
differently. The different renormalisations of the
various fields $\varphi^i$ would result in $N$ different
renormalised masses $m_i^2$ as well as the renormalised
couplings $\lambda^{ijkl}$ (the renormalised couplings are not all independent
parameters, of course, being functions of only
\hbox{${1\over 4!}N(N+1)(N+2)(N+3)+1$} bare couplings).
If however the bare theory enjoys global
$SO(N)$ invariance all the $\varphi_0^4$ couplings
reduce to only two, which
will be denoted by $\lambda_0$ and $\lambda_0^\prime$. Furthermore,
if there are no anomalies, this symmetry survives at the level
of the renormalised couplings to give only three renormalised parameters
$m^2,\lambda$ and $\lambda^\prime$. There is a Ward
identity which demands that all the renormalised masses $m_i^2$
must renormalise the same way,
i.e. $m_i^2=z_mm_0^2$, with the same renormalisation constant
$z_m$ for all the masses $m_i^2$. Similarly all of the renormalised
$\varphi^4$ couplings (a priori
${1\over 4!}N(N+1)(N+2)(N+3)$ in number) reduce to only two,
$\lambda$ and $\lambda^\prime$.
Thus the phase space, which is in principle
${2\over 4!}N(N+1)(N+2)(N+3)+2N$ dimensional, is reduced
to being only six dimensional by the symmetry.

A less trivial example is supplied by massless QED coupled
to a massless charged scalar field, with Lagrangian
$${\cal L}_0={1\over 4}F_0^2 +
i\overline\psi_0 \gamma^\mu D_\mu\psi_0 +
(\tilde D_\mu\varphi_0)^\dagger (\tilde D^\mu\varphi_0)
+{\lambda_0\over 4!}\varphi_0^4,\autoeq$$
\newcount\QEDL\QEDL=\count90
where the co-variant derivatives are defined by
$$D_\mu\psi_0
=(\partial_\mu +ie_0A_{0\mu})\psi_0\autoeq$$
$$\tilde D_\mu\varphi_0=(\partial_\mu +i\tilde e_0A_{0\mu})\varphi_0,
\autoeq$$
and
$$F_0^2=(\partial_\mu A_{0\nu}-\partial_\nu A_{0\mu})
(\partial^\mu A_0^\nu-\partial^\nu A_0^\mu).\autoeq$$
A coupling for the identity
operator is not necessary since the
theory is massless.
Also no gauge
fixing term is included for the moment because
a perturbative analysis will not be used here.
In order to avoid volume divergences
the theory can be formulated with periodic boundary conditions,
i.e. on a four dimensional torus
$T^4$.

There are three independent couplings, $e_0$, $\tilde e_0$
and $\lambda_0$, but Ward
identities force $e_0$ and  $\tilde e_0$
to renormalise in the same way so that
their renormalised couplings are related to the bare
counterparts
with the same renormalisation constant
(i.e. $e=ze_0$ and $\tilde e=z\tilde e_0$ where $z$ is the photon
wave function renormalisation constant $A_0^\mu=zA^\mu$).
Thus
$$\beta^e={de\over dt}=\left(z^{-1}{dz\over dt}\right)e
\qquad\hbox{and}\qquad
\beta^{\tilde e}={d\tilde e\over dt}=\left(z^{-1}{dz\over dt}\right)
\tilde e,\autoeq$$
which implies that $\tilde e\beta^e=\e\beta^{\tilde e}$.
The $\beta$-functions are therefore not independent. In particular
$${d\over dt}\left({\tilde e\over e}\right)=0,\autoeq$$
which immediately suggests a change of co-ordinates
from $(e,\tilde e)$ to $r=\sqrt{e^2+\tilde{e}^2}$ and
$\vartheta=\tan^{-1}(\tilde{e}/e)$, with $0\le\vartheta\le\pi/2$,
so that $\vartheta$ is
a RG invariant, ${d\vartheta\over dt}=0$.

The conjugate variables to $e$, $\tilde{e}$ and $\lambda$ are
$$\eqalign{
\phi_e&={i\over V}\int_{T^4}d^Dx<\overline\psi A^\mu\gamma_\mu\psi>\cr
\phi_{\tilde e}&={i\over V}\int_{T^4}d^Dx<A^\mu \tilde D_\mu\varphi>
\quad+\quad\hbox{Hermitian conjugate}\cr
\phi_\lambda&={1\over 4!V}\int_{T^4}d^Dx<\varphi^4>,\cr}
\autoeq$$
where $V$ is the volume of the torus. The integration over space
is kept explicit here so as to mantain gauge invariance
which is easily
proven by integrating by parts and using the equations of motion.
The two VEV's still have canonical mass dimension $D$ due to the
volume factors outside the integrals.

Since the theory is massless the potential $U$ vanishes and
the RG equations for the VEV's are
$$\eqalign{
{d\phi_e\over dt}+(\partial_e\beta^e)\phi_e
+(\partial_e\beta^{\tilde e})\phi_{\tilde e}
+(\partial_e\beta^\lambda)\phi_\lambda&=0\cr
{d\phi_{\tilde e}\over dt}+(\partial_{\tilde e}\beta^e)\phi_e
+(\partial_{\tilde e}\beta^{\tilde e})\phi_{\tilde e}
+(\partial_{\tilde e}\beta^\lambda)\phi_\lambda&=0\cr
{d\phi_\lambda\over dt}+(\partial_\lambda\beta^e)\phi_e
+(\partial_\lambda\beta^{\tilde e})\phi_{\tilde e}
+(\partial_\lambda\beta^\lambda)\phi_\lambda&=0.\cr}
\autoeq$$
However these are more elegantly expressed in the $(r,\vartheta)$
variables with
$$\eqalign{
\phi_r&=\cos\vartheta\;\phi_e +\sin\vartheta\;\phi_{\tilde e}\cr
\phi_\vartheta&=-r\sin\vartheta\;\phi_e + r\cos\vartheta\;\phi_{\tilde e}\cr}
\autoeq$$
and
$$\eqalign{
\beta^r&=\cos\vartheta\;\beta^e + \sin\vartheta\;\beta^{\tilde e}
={\beta^e\over\cos\vartheta}\cr
\beta^\vartheta&=0.\cr}\autoeq$$
Note that since $\cos\vartheta\ge0$ a positive $\beta^e$ always
gives a positive $\beta^r$, in agreement with the
statement that only a non-abelian theory can be
asymptotically free.

The RG evolution of the VEV's in terms of these new variables
is now given through
Hamilton's equations as
$$\eqalign{
{d\phi_r\over dt}+(\partial_r\beta^r)\phi_r
+(\partial_r\beta^\lambda)\phi_\lambda&=0\cr
{d\phi_\vartheta\over dt}+(\partial_\vartheta\beta^r)\phi_r
+(\partial_\vartheta\beta^\lambda)\phi_\lambda&=0\cr
{d\phi_\lambda\over dt}+(\partial_\lambda\beta^r)\phi_r
+(\partial_\lambda\beta^\lambda)\phi_\lambda&=0.\cr}
\autoeq$$
\newcount\dphit\dphit=\count90

Since $\beta^\vartheta$ vanishes the Hamiltonian
$$H(r,\vartheta,\lambda,\phi_r\phi_\lambda)=
\beta^r(r,\vartheta,\lambda)\phi_r
+\beta^\lambda(r,\vartheta\lambda)\phi_\lambda\autoeq$$
\newcount\QEDH\QEDH=\count90
is independent of $\phi_\vartheta$ and
$\vartheta$ is a constant of the motion. In analogy with classical
mechanics $\vartheta$ might be called an ignorable co-ordinate,
but the roles of co-ordinate and momenta in the RG are
really reversed from those of classical mechanics.
It is more correct to say the $\phi_\vartheta$ is an ignorable  expectation
value because the Hamiltonian is still a non-trivial function of $\vartheta$.
The situation is in fact more involved here than in classical mechanics
since the momentum dual to an ignorable co-ordinate in classical mechanics
only appears quadratically in the Hamiltonian, whereas the
$\vartheta$ dependence of $H$ in equation (\the\QEDH) can be much
more complicated. The invariant $\vartheta$ however still plays
the same role as an invariant in classical mechanics -
since $\{\vartheta,H\}=0$ it generates a symmetry on phase space
via the Poisson bracket operation. This can be viewed as the
implementation of the Ward identities on phase space.


The fact that one of the expectation values can be eliminated
from the phase space can be understood from a physical point
of view in the following manner. If the gauge field $A_{0\mu}$
is rescaled by $A_{0\mu}\rightarrow{1\over e_0}A_{0\mu}$ then
the gauge coupling completely drops out of the matter field terms
in the Lagrangian (\the\QEDL) and only appears in the kinetic energy term
for the gauge fields, ${1\over 4e_0^2}F_0^2$, only the ratio
$\vartheta$ appears in the matter field Lagrangian and this is a
RG invariant. Defining a new variable $q_0={1\over e_0^2}$
and forgetting about $\vartheta$ there are now only two
expectation values to be considered, $\phi_q={1\over 4}<F^2>$
and $\phi_\lambda$.
One may expect $\phi_q\ne 0$ at large energies, even in
a massless theory, since the
$\beta$-function for $e$ is
positive
\autoref\newcount\Miransky\Miransky=\refno.

There are further interesting aspects of gauge theories
when a gauge fixing term is added. If a term
$\eta_0(\partial_\mu A^\mu_0)^2$ is introduced into
the Lagrangian (\the\QEDL) then there is another
$\beta$-function, $\beta^\eta={d\eta\over dt}$.
In minimal subtraction schemes, the $\beta$-functions
for the other couplings are independent of the gauge fixing parameter $\eta$,
so
$\partial_\eta\beta^r(r,\vartheta\lambda)=
\partial_\eta\beta^\lambda(r,\vartheta\lambda)=0$.
Therefore the expectation value
$\phi_\eta={1\over V}\int_{T^4}d^Dx<(\partial.A)^2>$
evolves under RG flow according to
$${d\phi_\eta\over dt}+(\partial_\eta\beta^\eta)\phi_\eta=0
\qquad\Rightarrow\qquad \phi_\eta(t)=\phi_\eta(t_0)
\e^{-\int_{t_0}^t(\partial_\eta\beta^\eta)dt}.\autoeq$$
If $\phi_\eta$ is zero at some value of $t$, then it is zero
at all values and
$\phi_\eta$ is another constant constant of the motion.
This reflects the fact that $\eta$ plays no
physical role in the theory. If a renormalisation
prescription other than minimal
subtraction is used, however, it may not be the case
that the other couplings have $\beta$-functions
which are independent of the gauge fixing
parameter. However, holding to the philosophy that a change in
regularisation
prescription is just a change in co-ordinates, there must be {\it some}
quantity that is a RG invariant, i.e. the co-ordinate transform
of the dimensional regularisation co-ordinate $\eta$ -
it may look messy in the new co-ordinates but it must
exist.

In conclusion it would seem that, just as in classical mechanics,
the Hamiltonian framework is a very powerful one for the discussion
of symmetries, which play such a central role in all discussions
of quantum field theory.
\vfill\eject
{\bf \S 6 Conclusions}
\vskip .5cm
Before summarising the main results of this paper
a few comments will be made about the global topology of the space of
couplings $\widehat{\cal M}$, as promised in the introduction.
Consider for definiteness massless QED with the only coupling
being the electron charge $e$ (the identity
operator can be ignored). It is really $\alpha={e^2\over\hbar c}$
(or $1/\alpha$) which is the important parameter, and $\alpha$ must be positive
since a negative value would mean that the theory would be
unstable, as pointed out by Dyson
\autoref\newcount\Dyson\Dyson=\refno.
Thus $\alpha=0$ is not an analytic point and one
cannot continue, even infinitesimally, to $\alpha<0$. Similarly
$1/\alpha=0$ cannot be an analytic point. The manifold $\cal M$,
the positive real line, has two boundary points both of which are
non-analytic points of the theory. For the higher dimensional case
points of non-analyticity are clearly also of central importance.
Such points might be isolated or might form sub-manifolds of
$\widehat{\cal M}$ with dimension $k<n$. If the non-analytic points were
to form a sub-manifold of co-dimension one (a hypersurface
$k=n-1$), then this would act as an effective boundary
$\partial\widehat{\cal M}$. In any event, it is clear that
any understanding of the global topology of $\widehat{\cal M}$
will be inextricably linked with an understanding of the points
of non-analyticity.

In summary it has been argued that
the renormalisation group evolution
of couplings and vacuum expectation values can be described
as a Hamiltonian flow on the $2n$ dimensional
phase space $T^*(\widehat{\cal M})$ with the Hamiltonian
given by
$$H(g,\phi,t)=\beta^a(g,t)\phi_a +
\bigl(U(g,t)-D\Lambda\bigr)\phi_\Lambda,\autoeq$$
\newcount\conham\conham=\refno
which can be identified with minus
the vacuum expectation value of the trace of the energy-momentum
tensor.

The natural variables canonically
conjugate to the couplings are the expectation values,
$\phi_a=\partial_a w(g,t)$, where $w(g,t)$ is the generating
functional or free energy density.
For theories with massive couplings
the cosmological constant plays a central role, since its
$\beta$-function $\beta_\Lambda=U(g,t)-D\Lambda$ gives rise to a
potential which acts as an effective force in the RG evolution
of the VEV's
$${d\phi\over dt}=-\kappa^DdU(g,t),\autoeq$$
where $\phi=\phi_adg^a$ and $dU=\partial_aUdg^a$.

The RG evolution of any function $A(g,\phi,t)$ can be determined
from the Hamiltonian
$${dA\over dt}
=\{A,H\}+{\partial A\over\partial t}\Vline\sub{g,\phi}.
\autoeq$$
In particular the RG evolution of the Hamiltonian itself is
given by
$${dH\over dt}={\partial H\over\partial t}=\partial_t\beta^a(g,t)\phi_a
+\partial_tU(g,t)\phi_\Lambda,\autoeq$$
and $H$ is a RG invariant if a subtraction procedure is chosen
so that the $\beta$-functions are independent of $\kappa$.
Alternatively the cosmological constant can be omitted and
the RG evolution of the Hamiltonian $h$ on $T^*({\cal M})$ is given
by
$${dh\over dt}={\partial h\over\partial t}=\partial_t\beta^a(g,t)\phi_a
+\kappa^D\bigl(DU(g,t)+\partial_tU(g,t)\bigr),\autoeq$$
which can only be a RG invariant if both $U=0$
and the $\beta$-functions have no explicit $\kappa$
dependence.

The RG equation for the $N$-point Green functions is
$$\eqalign{
{\partial\over\partial t}G^{(N)}_{a_1\cdots a_N}(g,\phi,t)&
+\beta^b\partial_bG^{(N)}_{a_1\cdots a_N}(g,\phi,t)
+\sum_{i=1}^N(\partial_{a_i}\beta^b)
G^{(N)}_{a_1\cdots a_{i-1}ba_{i+1}\cdots a_N}(g,\phi,t)\cr
&=\Bigl(\phi_c(\partial_b\beta^c)
+\kappa^D\partial_b U\Bigr){\partial\over\partial\phi_b}
G^{(N)}_{a_1\cdots a_N}(g,\phi,t).\cr}\eqno(\the\RGN)$$
The crucial ingredient that gives rise to Hamiltonian flow
is the underlying symmetry of the renormalisation group,
reflected in the fact that ${dw\over dt}=0$.
Ward identities give rise to further constants of the
motion which generate symmetries on phase space via
the Poisson bracket structure.

The RG equation for the generating functional $w(g(t),t)$ can be interpreted
as a Hamilton-Jacobi equation
$${\partial w\over\partial t}\Vline\sub g +
H(g,{\partial w\over\partial g},t)=0.\eqno(62)$$

Table $1$ provides a summary of the correspondence
between concepts in quantum field theory or statistical mechanics
and classical mechanics

It is a pleasure to thank William Deans, Chris Ford, Bill McGlinn,
Charles Nash, and Lochlainn O'Raifeartaigh for many useful
conversations about the RG. It was Denjoe O'Connor and Chris Stephens
who first suggested to me that a symplectic structure should play
an important role in quantum field theory and statistical
mechanics, with the VEV's being conjugate momenta, and that the RG
equation should be viewed as a Hamilton-Jacobi like equation.
This work would not have been possible without their prompting.
I also wish to thank the Alexander von Humboldt foundation
whose financial support enabled me to carry out most of the calculations
of section four.
\vfill\eject
{\bf References}\hfil
\vskip .5cm
\item{[\the\Brezin]} E. Br\'ezin, J.C. Le Guillou and J. Zinn-Justin,
Phys. Lett. {\bf 44A} (1973) 227;\hfil\break
Phys. Rev. B {\bf 10} (1973) 892
\smallskip
\item{[\the\DenjoeChris]} {\it Geometry, The Renormalisation Group
And Gravity}
D. O'Connor and C.R. Stephens, \break
in {\it Directions In General Relativity},\hfill\break
Ed. B.L. Hu, M.P. Ryan Jr., and C.V. Vishveshwava\hfill\break
Proceedings of the 1993 International Symposium,
Maryland, Vol 1, C.U.P. (1993)
\smallskip
\item{[\the\BC]} L.S. Brown, Ann. Phys. {\bf 126} (1979) 135\hfil\break
L.S. Brown and J.C. Collins, Ann. Phys. {\bf 130} (1980) 215
\smallskip
\item{[\the\WZia]} D.J. Wallace and R.K.P. Zia, Phys. Lett. {\bf A48} (1974)
565\hfil\break
D.J. Wallace and R.K.P. Zia, Ann. Phys. {\bf 92} (1975) 142
\smallskip
\item{[\the\Vasiliev]} A.N. Vasil'ev, A.K. Kazanskii and Yu. M. Pis'mak,
Teor. i Math. Fiz. {\bf 19} (1974) 186\hfil\break
J.M. Cornwall, R. Jackiw and E. Tomboulis, Phys. Rev D {\bf 10} (1974)
2428\hfil\break
G.M. Shore, Nucl. Phys. {\bf B362}, (1991), 85
\smallskip
\item{[\the\Zam]} A.B. Zamolodchikov, Rev. Math. Phys. {\bf 1}, (1990), 197
\smallskip
\item{[\the\ChengLi]}
Ta-Pei Cheng and Ling-Fong Li, {\it Gauge Theory Of Elementary Particle
Physics}\hfil\break
(1984) O.U.P.
\smallskip
\item{[\the\Tod]} B.P. Dolan, Int. J. of Mod. Phys. A {\bf 9} (1994) 1261
\smallskip
\item{[\the\Arnold]}
V.I. Arnold, {\it Mathematical Methods Of Classical Mechanics} 2nd.
Ed.\hfil\break
Graduate Texts in Mathematics {\bf 60}, (1989) Springer
\smallskip
\item{[\the\Lassig]} M. L\"assig, Nucl. Phys. {\bf B334}, (1990), 652
\smallskip
\item{[\the\DCup]} D. O'Connor and C.R. Stephens, unpublished
\smallskip
\item{[\the\HughIan]} I. Jack and H. Osborn, Nucl. Phys. {\bf B343}, (1990),
647
\smallskip
\item{[\the\JL]} G. Jona-Lasinio, Nuovo Cimento {\bf 34} (1964) 1790
\smallskip
\item{[\the\Hathrell]} S.J. Hathrell, Ann. Phys. {\bf 139} (1982) 136
\smallskip
\item{[\the\CW]} C.G. Callan Jr., S. Coleman and R. Jackiw,
Ann. Phys. {\bf 59} (1970) 42
\smallskip
\item{[\the\Miransky]} V.A. Miransky, Il Nouvo Cimento {\bf 90} (1985) 149
\smallskip
\item{[\the\Dyson]} F.J. Dyson, Phys. Rev. {\bf 85} (1952) 631
\smallskip
\vfill\eject
\nopagenumbers
\centerline{A Quantum Field Theory-Classical Mechanics Dictionary}
\vskip .5cm
\vbox{\offinterlineskip
\hrule
\halign{&\vrule#&
  \strut\quad#\hfil\quad\cr
height 2pt&\omit&&\omit&\cr
&Quantum Field Theory or\hfil&&Classical\ &\cr
height 2pt&\omit&&\omit&\cr
&Statistical Mechanics\hfil&&Mechanics&\cr
height 2pt&\omit&&\omit&\cr
\noalign{\hrule}
height 2pt&\omit&&\omit&\cr
&Couplings $\qquad g^a(t)$&&Co-ordinates $\qquad q^a(t)$&\cr
height 2pt&\omit&&\omit&\cr
&$\beta$-functions $\qquad\beta^a(t)$&&Velocities $\qquad\dot q^a(t)$&\cr
height 2pt&\omit&&\omit&\cr
&Vacuum expectation values $\quad\phi_a(t)$&&Momenta $\qquad p_a(t)$&\cr
height 2pt&\omit&&\omit&\cr
&Bare couplings $\qquad (g_0^a,\phi^0_a)$ &&Initial point $\qquad
(q_0^a,p_a^0)$&\cr
height 2pt&\omit&&\omit&\cr
&&&&\cr
height 2pt&\omit&&\omit&\cr
&Generating functional $\quad w(g(t),g_0,t)$&&Action $\qquad S(q(t),q_0,t)$
&\cr
height 2pt&\omit&&\omit&\cr
&(Free energy density) &&(Hamilton's principal function)&\cr
height 2pt&\omit&&\omit&\cr
&&&&\cr
height 2pt&\omit&&\omit&\cr
&Hamiltonian&&Hamiltonian&\cr
height 2pt&\omit&&\omit&\cr
&$H(g,\phi,t)=\beta^{\hat a}(g,t)\phi_{\hat a}$
&&$H(q,p,t)={1\over 2m}g^{ab}(q)p_ap_b+U(q,t)$ &\cr
height 2pt&\omit&&\omit&\cr
&Hamilton's equations &&Hamilton's equations&\cr
height 2pt&\omit&&\omit&\cr
&$\beta^a={\partial H\over\partial\phi_a}\qquad
\dot\phi_a=-{\partial H\over\partial g^a}$
&&$\dot q^a={\partial H\over\partial p_a}\qquad
\dot p_a=-{\partial H\over\partial q^a}$&\cr
height 2pt&\omit&&\omit&\cr
&Potential $\qquad U(g,t)=\beta^\Lambda +D\Lambda$
&&Potential $\qquad U(q,t)$&\cr
height 2pt&\omit&&\omit&\cr
&RG flow of VEV's $\qquad{d\phi\over dt}=-dU$
&&Newton's $2^{\hbox{nd}}$ law $\quad{dp\over dt}=-dU$&\cr
height 2pt&\omit&&\omit&\cr
& && &\cr
height 2pt&\omit&&\omit&\cr
&RG equation for $w$&&Hamilton-Jacobi equation &\cr
height 2pt&\omit&&\omit&\cr
&${\partial w\over\partial t}+
H\Bigl(g(t),{\partial w\over\partial g},t\Bigr)=0$
&&${\partial S\over\partial t}+
H\Bigl(q(t),{\partial S\over\partial q},t\Bigr)=0$ &\cr
height 2pt&\omit&&\omit&\cr
&&&&\cr
height 2pt&\omit&&\omit&\cr
&Massless theory $\qquad U=0$&&Free particle $\qquad U=0$&\cr
height 2pt&\omit&&\omit&\cr
&No explicit $\kappa$ dependence in $\beta$&&Conservative system&\cr
height 2pt&\omit&&\omit&\cr
&Anomalous dimensions &&Pseudo-forces (Coriolis) &\cr
height 2pt&\omit&&\omit&\cr
&RG invariant $\qquad\{\vartheta,H\}=0$
&&Constant of motion $\quad\{\vartheta,H\}=0$&\cr
height 2pt&\omit&&\omit&\cr
&(Ward identity)&&(Symmetry generator)&\cr
height 2pt&\omit&&\omit&\cr}
\hrule}
\bye